\documentclass[12pt,a4paper]{article}
\usepackage[english]{babel}
\usepackage[latin1]{inputenc}
\usepackage[T1]{fontenc}
\usepackage{amsmath}
\usepackage{amsfonts}
\usepackage{amssymb}
\usepackage[dvips]{graphicx,color}
\begin{document}
\title{Vacuum polarization from confined fermions in $3+1$ dimensions} 
\author{C. Ccapa Ttira, C.~D.~Fosco and
E.~Losada\\
{\normalsize\it Centro At\'omico Bariloche and Instituto Balseiro}\\
{\normalsize\it Comisi\'on Nacional de Energ\'{\i}a At\'omica}\\
{\normalsize\it R8402AGP S.\ C.\ de Bariloche, Argentina.} }
\maketitle
\begin{abstract} 
We study the main properties of the one-loop vacuum polarization function
($\Pi_{\mu \nu}$) for massless spinor $QED_4$ in a slab, namely, with
fields defined on ${\mathcal M} \subset {\mathbb R}^{3+1}$, such that
${\mathcal M} = \{(x_0,...,x_3) | 0 \leq x_3 \leq  \epsilon\}$, and
bag-like boundary conditions on the boundary.  We evaluate the induced
charge density and current due to an external constant electric field
normal to the boundary; we also study the effective action for a purely
transverse field, identifying its $\epsilon$-dependent contribution.
\end{abstract}
\maketitle
In the presence of nontrivial boundary conditions, Quantum Field Theory
models may give rise to many interesting effects.  Noteworthy examples of them
are the Casimir effect~\cite{Casimir:1948dh,Bordag:2001qi}, the bag model of
QCD~\cite{Chodos:1974pn}, as well as many others~\cite{Milton:2001yy}.

The common origin of those effects is the presence of boundaries, which
strongly affect the structure of the vacuum fluctuations; this fact has
relevance not only for global observables, i.e., Casimir energies, but also
for local ones, like vacuum energy densities and response functions. The
latter, which are determined by the correlation between fluctuations should
exhibit a strong dependence with the distance to the boundary (at least for
the ones involving degrees of freedom affected by the boundary conditions). 

In this letter, we consider the vacuum polarization tensor, $\Pi_{\mu\nu}$,
for a Dirac field confined to a slab-shaped region ${\mathcal M}$, defined
by the condition $0 \leq x_3 \leq \epsilon$.  More concretely, we consider
a massless Dirac field in $3+1$ dimensions, which satisfies `bag' (i.e.,
vanishing normal current) boundary conditions on the two static planes
$x_3=0$ and $x_3=\epsilon$. Euclidean spacetime coordinates are, in the
conventions that we shall use, denoted by $(x_0,x_1,x_2,x_3)$.  Moreover,
we shall assume the Dirac field to be confined to the region between those
mirrors; hence, $\Pi_{\mu\nu}$ vanishes identically outside of $0 < x_3 <
\epsilon$. It there are fermions outside of the slab, $\Pi_{\mu\nu}$ does
not vanish there. However, their properties in that region are equivalent
to the case of just one boundary; for example, if $x_3 < 0$,  $\Pi_{\mu\nu}$ 
in that region is the same as the one for a single wall at $x_3=0$.

\noindent In the conventions we shall use, Euclidean coordinates are
denoted by $x_\mu$, $\mu= 0,1,2,3$, while the metric tensor is given by
$g_{\mu\nu}=\delta_{\mu\nu}$.

To account for the effect of the confined fermion fluctuations on the gauge
field propagation, one introduces:
\begin{equation}\label{eq:defgef}
	e^{- \Gamma_f(A)} \;=\; \frac{\int {\mathcal D}\psi {\mathcal
	D}{\bar\psi} \, e^{-{\mathcal S}_f({\bar\psi},\psi;A)}}{\int 
	{\mathcal D}\psi {\mathcal D}{\bar\psi} \, 
	e^{-{\mathcal S}_f({\bar\psi},\psi;0)}} \;,
\end{equation}
where the fermionic action, $S_f$ accounts for the (minimal) coupling to
the gauge field as well as for the introduction of the bag boundary
conditions. To deal with the latter, we follow the approach of representing 
them by local interaction terms~\cite{EfeCasFer}. Of course,
the resulting propagator should agree with the one one would get by using,
for example, the multiple reflection expansion (MRE)
\cite{Balian:1970fw,Hansson:1982cu,Hansson:1983xt}. That approach has been
used in~\cite{Fosco:2004cn} for the calculation of $\Pi_{\mu\nu}$ for a
Dirac field in a half-space.  

Following~\cite{EfeCasFer}, we include a `potential' $V$ into the fermionic
action, so that:   
\begin{equation}
	{\mathcal S}_f({\bar\psi},\psi;A) \;=\; \int d^4x \, {\bar\psi}(x) 
\big[	\not\! \partial + g \, V(x_3) + i e \not \!\! A(x) \big] \psi(x) 
\end{equation}
with
\begin{equation}
V(x_3) \;=\; \big[ \delta(x_3) \,+\, \delta(x_3 - \epsilon) \big] \;.
\end{equation}
Here, $g$ is a constant which, in order to enforce bag boundary conditions,
has to be equal to $2$~\footnote{See~\cite{EfeCasFer}. Different values would produce `imperfect'
boundary conditions}. The slash notation denotes contraction with the
Euclidean $\gamma$-matrices which, in our conventions, are all Hermitian,
and satisfy:
\mbox{$\{\gamma_\mu \, ,\, \gamma_\mu \} = 2 \, \delta_{\mu\nu}$},
$\mu,\, \nu\, =\, 0,\, 1,\, 2,\, 3$. 

From (\ref{eq:defgef}), we may write,
\begin{equation}\label{eq:gef}
	\Gamma_f(A) \;=\; - {\rm Tr} \, \ln \left[ 1 \,+\, i e \, 
	(\not\!\partial + 2 V)^{-1} \not \!\! A \right] \;,
\end{equation}
which allows us to introduce the vacuum polarization, $\Pi_{\mu\nu}$, the
kernel determining the form of the first non-trivial term for the expansion
of $\Gamma_f$ in powers of $A$:
\begin{equation}
	\Gamma_f(A) \;=\; \frac{1}{2} \,\int d^4x \int d^4y \, 
	A_\mu(x) \, \Pi_{\mu\nu}(x,y) \, A_\nu(y) \;+\;\ldots
\end{equation}

From (\ref{eq:gef}), we may give a more explicit, yet formal, 
expression for $\Pi_{\mu\nu}$ in coordinate space:
\begin{equation}\label{eq:pimn}	
	\Pi_{\mu\nu}(x,y) \;=\; - e^2 \, {\rm tr} \left[ 
	S_F(y,x) \gamma_\mu S_F(x,y) \gamma_\nu \right] \;,
\end{equation}
where $S_F$ denotes the (coordinate space) exact fermion propagator for $e=0$, namely, the
free propagator with bag boundary conditions.
This object may be obtained as the matrix elements of the inverse of an
operator:
\begin{equation}\label{eq:defsf}
	S_F(x,y) \;=\; \langle x|(\not\!\partial +  2 \, V)^{-1} |y\rangle \;,
\end{equation}
where a `bracket' notation has been used to denote the coordinate space
versions of operators (in this case, the inverse of $\not\!\!\partial + 2
V$).  Since the boundary conditions only affect the $x_3$ coordinate, 
there is symmetry under translations in the `parallel' coordinates
$x_\shortparallel \equiv (x_0,x_1,x_2)$. Thus, $S_F =
S_F(x_\shortparallel-y_\shortparallel;x_3,y_3)$, and it is natural to work
with $\widetilde{S}_F$, a mixed Fourier representation of $S_F$ which
results by transforming the parallel coordinates only, so that 
\mbox{$S_F(x_\shortparallel-y_\shortparallel;x_3,y_3) \to
\widetilde{S}_F(p_\shortparallel;x_3,y_3)$}. 

This reduces the problem to a one-dimensional one, where $\widetilde{S}_F$
satisfies the equation:
\begin{equation}\label{eq:dif} 
\big[ \gamma _3\,\partial _{x_3}  - i \not \! p_\shortparallel + 2 V(x_3)\Big]\,
\widetilde{S}_F(p_\shortparallel ; x_3 ,y_3)\,=\, \delta (x_3 - y_3)\;.  
\end{equation} 
The solution to the inhomogeneous linear equation above may be written as the sum of
two terms: 
\begin{equation}\label{eq:suma} 
\widetilde{S}_F(p_\shortparallel ; x_3 ,y_3) \,=\, 
\widetilde{S}_F^{(0)}(p_\shortparallel ; x_3 ,y_3) \,+\, \widetilde{U}(p_\shortparallel ; x_3 ,y_3)\;,
\end{equation}
where the first of them, $\widetilde{S}_F^{(0)}$, is the free fermion propagator in the absence of boundaries,
while the other, denoted by $U$ accounts for the boundary (bag) conditions. 

The explicit form of $\widetilde{S}_F^{(0)}$ may be found quite straightforwardly:
\begin{equation}\label{eq:s2} 
	\widetilde{S}_F^{(0)}(p_\shortparallel ; x_3 ,y_3)
	\,=\,
	\frac{1}{2}\big[ \mathrm{sgn}(x_3 - y_3)\,\gamma _3 - i \frac{\not \! p_\shortparallel}{\vert
	p_\shortparallel \vert}\big]\;e^{- \vert p_\shortparallel \vert
	\vert x_3 - y_3\vert }\;.  
\end{equation}

Since $\widetilde{S}_F^{(0)}$ is the Green's
function of the free Dirac operator we derive, from (\ref{eq:dif}), an
equation for $\widetilde{U}(p_\shortparallel , x_3 ,y_3)$:
\begin{equation} 
\big( \gamma _3\,\partial _{x_3}  + i \not \!  p_\shortparallel \big)\,
\widetilde{U}(p_\shortparallel ; x_3 ,y_3)
\,=\,-\, 2 \, V(x_3) \,\big[\widetilde{S}_F^{(0)}(p_\shortparallel ; x_3
,y_3) + \,\widetilde{U}(p_\shortparallel ; x_3 ,y_3)\big]\;, 
\end{equation}
which implies:
\begin{eqnarray}\label{eq:uu} 
	\widetilde{U}(p_\shortparallel ; x_3 ,y_3) &=&- 2\,\Big[
	\widetilde{S}_F^{(0)}(p_\shortparallel ; x_3 ,0)\,
	\widetilde{S}_F^{(0)}(p_\shortparallel ; 0
	,y_3)\,+\,\widetilde{S}_F^{(0)}(p_\shortparallel ; x_3
	,0)\,\widetilde{U}(p_\shortparallel ; 0 , y_3)
\nonumber\\ 
&+& \widetilde{S}_F^{(0)}(p_\shortparallel ; x_3 ,\epsilon)\,\widetilde{S}_F^{(0)}(p_\shortparallel ; \epsilon ,y_3)
\,+\,\widetilde{S}_F^{(0)}(p_\shortparallel ; x_3 ,\epsilon)\,\widetilde{U}(p_\shortparallel ,
\epsilon ,y_3) \Big]\;.  
\end{eqnarray}
One can then obtain $\widetilde{U}(p_\shortparallel ; x_3 ,y_3)$  from the equation
above, for example by first finding $\widetilde{U}(p_\shortparallel , 0 ,y_3)$ and
$\widetilde{U}(p_\shortparallel , \epsilon ,y_3)$: these can be found by evaluating
(\ref{eq:uu}) at $x_3 = 0$ and $x_3 = \epsilon$, respectively, and then
solving the resulting system of equations.

Once those objects are found, the outcome of this procedure is an explicit
expression for $\widetilde{U}(p_\shortparallel ;x_3,y_3)$, which can be expressed as a
sum of terms, distinguished according to its $\gamma$-matrix content:
\begin{eqnarray}\label{eq:propU}
	\widetilde{U}(p_\shortparallel;x_3,y_3) &=& \widetilde{U}_0(p_\shortparallel;x_3,y_3) I +
	\widetilde{U}_1(p_\shortparallel;x_3,y_3) \big(- i \frac{\not\! p
_\shortparallel}{\vert p _\shortparallel \vert}\big) \nonumber\\ 
&+& \widetilde{U}_2(p_\shortparallel;x_3,y_3) 
\big(- i \frac{\not\! p_\shortparallel}{\vert p_\shortparallel \vert} \gamma_3 \big) +
\widetilde{U}_3(p_\shortparallel,x_3,y_3) \gamma _3\;, 
\end{eqnarray}
where we have introduced four functions $\widetilde{U}_a$ ($a=0,1,2,3$). The
explicit form of these functions, for $0<x_3<\epsilon$ and
$0<y_3<\epsilon$, is the following:
\begin{eqnarray}\label{eq:defu0}
	\widetilde{U}_0(p_\shortparallel;x_3,y_3) &=& \frac{1}{2 \big(e^{2
	|p_\shortparallel| \epsilon} + 1\big)} \, \Big[ e^{|p_\shortparallel|
	(x_3 + y_3)} + e^{-|p_\shortparallel| (x_3 + y_3 - 2 \epsilon )}
	\Big] \;,
\end{eqnarray}
\begin{eqnarray}\label{eq:defu1}
	\widetilde{U}_1(p_\shortparallel;x_3,y_3) &=& - \frac{1}{2 \big(e^{2
	|p_\shortparallel| \epsilon} + 1\big)} \, 
\Big[ e^{|p_\shortparallel| (x_3 - y_3)} + e^{-|p_\shortparallel| (x_3 -
y_3)}\Big] \;,
\end{eqnarray}
\begin{eqnarray}\label{eq:defu2}
	\widetilde{U}_2(p_\shortparallel;x_3,y_3) &=& - \frac{1}{2 \big(e^{2 |p_\shortparallel| \epsilon} + 1\big)} \, 
\Big[ e^{|p_\shortparallel| (x_3 + y_3)} - e^{-|p_\shortparallel| (x_3 +
y_3 - 2 \epsilon)}\Big]
\end{eqnarray}
\begin{eqnarray}\label{eq:defu3}
	\widetilde{U}_3(p_\shortparallel;x_3,y_3) &=& \frac{1}{2 \big(e^{2 |p_\shortparallel| \epsilon} + 1\big)} \, 
\Big[ e^{|p_\shortparallel| (x_3 - y_3)} - e^{-|p_\shortparallel| (x_3 -
y_3)}\Big] \;.
\end{eqnarray}
A lengthy but otherwise straightforward calculation shows that the bag
boundary conditions are fulfilled, namely, the following equations are satisfied:
\begin{eqnarray}
\lim_{x_3 \to 0+} (I - \gamma_3) \widetilde{S}_F(p_\shortparallel;x_3,y_3)
&=& 0 \nonumber\\
\lim_{x_3 \to \epsilon-} (I + \gamma_3) \widetilde{S}_F(p_\shortparallel;x_3,y_3)
&=& 0 
\end{eqnarray} 
(as well as the ones corresponding to approaching the boundaries with the
$y_3$ coordinate).

We now analyze the UV properties of the propagator $\widetilde{S}_F$, as
presented in (\ref{eq:suma}). We see that the two terms have a quite
different behaviour. Indeed, the first one, $\widetilde{S}^{(0)}_F$, being
the free propagator in the absence of boundaries, does have the well-known
UV behaviour ($\sim p^{-1}$ in momentum space). In the `mixed' Fourier
representation, where it depends on $p_\shortparallel$ and $x_3,\, y_3$,
that behaviour translates into a $\sim |p_\shortparallel|^0$  behaviour
when $x_3=y_3$, and an exponential decay when $x_3 \neq y_3$.  The
$\widetilde{U}$ term, on the other hand, decays exponentially everywhere,
except when both $x_3$ and $y_3$ approach  one (the same) boundary, namely,
when either $x_3+y_3 \to 0$ or $x_3 + y_3 \to 2 \epsilon$. Moreover, we can
trace the origin of that behaviour to see that it comes from the
$\widetilde{U}_0$ and $\widetilde{U}_2$ terms ($\widetilde{U}_1$ and
$\widetilde{U}_3$ always decrease exponentially, regardless of the values
of $x_3$ and $y_3$).

As a consistency check for the expression of $S_F$ , one can find an alternative representation,
obtained by expressing $\big(e^{2 |p_\shortparallel| \epsilon} +
1\big)^{-1}$, which appears as a common factor in the expressions for
$\widetilde{U}_a$, as a series:
\begin{equation}
\frac{1}{e^{2 |p_\shortparallel| \epsilon} + 1} = 
\sum_{n=0}^\infty \, e^{ - 2 (n+1) |p_\shortparallel| \epsilon} \;.
\end{equation} 
As a result, and after some algebra, we obtain:
\begin{eqnarray}
\widetilde{S}_F(p_\shortparallel;x_3,y_3) &=& \sum_{n=-\infty}^{+\infty}
(-1)^n \, \Big[ \widetilde{S}^{(0)}_F(p_\shortparallel;x_3,y_3 + 2 n \epsilon)
\nonumber\\
&+& \, \gamma_3 \; \widetilde{S}^{(0)}_F(p_\shortparallel;x_3,-y_3 + 2 n \epsilon)
\Big] \;,
\end{eqnarray}
which may naturally be thought of as an MRE representation of $\widetilde{S}_F$. 
It is possible to pinpoint also here the part of the propagator that
control its UV behaviour.  Since all the terms in the sum can be written as
functions of $\widetilde{S}^{(0)}_F$, it is quite straightforward to see
that the terms that control the large-$|p_\shortparallel|$ regime are:
$n=0$ (for $x_3=y_3$ or $x_3+y_3=0$) and $n=1$ (for $x_3+y_3=2 \epsilon$).

To obtain a more explicit expression for $\Pi_{\mu\nu}$, we insert
the results derived previously for $S_F$ into (\ref{eq:pimn}). We first
note that, since $\Pi_{\mu\nu}$ shall also be a function of
\mbox{$(x_\shortparallel-y_\shortparallel; x_3, y_3)$}, we may write its
Fourier transform with respect to the parallel variables, as follows: 
\begin{equation}\label{eq:pimn1}	
	\widetilde{\Pi}_{\mu\nu}(k_\shortparallel; x_3,y_3) \;=\; - e^2 \, \int
	\frac{d^3p_\shortparallel}{(2\pi)^3} \; {\rm tr} \left[ 
	\widetilde{S}_F(p_\shortparallel; y_3,x_3) \gamma_\mu
	\widetilde{S}_F(p_\shortparallel + k_\shortparallel; x_3,y_3) \gamma_\nu \right] \;.
\end{equation}
Thus,
\begin{eqnarray}\label{eq:pimn2}	
	\widetilde{\Pi}_{\mu\nu}(k_\shortparallel; x_3,y_3) &=& 
	\widetilde{\Pi}^{LL}_{\mu\nu}(k_\shortparallel; x_3,y_3) \,+\, 
	\widetilde{\Pi}^{LU}_{\mu\nu}(k_\shortparallel; x_3,y_3) \nonumber\\
	&+&\widetilde{\Pi}^{UL}_{\mu\nu}(k_\shortparallel; x_3,y_3) \,+\, 
	\widetilde{\Pi}^{UU}_{\mu\nu}(k_\shortparallel; x_3,y_3) \;, 
\end{eqnarray}
where:
\begin{equation}\label{eq:pill}	
	\widetilde{\Pi}^{LL}_{\mu\nu}(k_\shortparallel; x_3,y_3) = - e^2 \int
	\frac{d^3p_\shortparallel}{(2\pi)^3} {\rm tr} \left[ 
	\widetilde{S}_F^{(0)}(p_\shortparallel; y_3,x_3) \gamma_\mu
	\widetilde{S}_F^{(0)}(p_\shortparallel + k_\shortparallel; x_3,y_3) \gamma_\nu \right] \;,
\end{equation}
\begin{equation}\label{eq:pilu}	
	\widetilde{\Pi}^{LU}_{\mu\nu}(k_\shortparallel; x_3,y_3) = - e^2 \int
	\frac{d^3p_\shortparallel}{(2\pi)^3} \; {\rm tr} \left[ 
	\widetilde{S}_F^{(0)}(p_\shortparallel; y_3,x_3) \gamma_\mu
	\widetilde{U}(p_\shortparallel + k_\shortparallel; x_3,y_3)
	\gamma_\nu \right] \;,
\end{equation}
\begin{equation}\label{eq:piul}	
	\widetilde{\Pi}^{UL}_{\mu\nu}(k_\shortparallel; x_3,y_3)=- e^2 \, \int
	\frac{d^3p_\shortparallel}{(2\pi)^3} {\rm tr} \left[ 
	\widetilde{U}(p_\shortparallel; y_3,x_3) \gamma_\mu
	\widetilde{S}_F^{(0)}(p_\shortparallel + k_\shortparallel; x_3,y_3) \gamma_\nu \right] \;,
\end{equation}
\begin{equation}\label{eq:piuu}	
	\widetilde{\Pi}^{UU}_{\mu\nu}(k_\shortparallel; x_3,y_3) =- e^2 \int
	\frac{d^3p_\shortparallel}{(2\pi)^3} {\rm tr} \left[ 
	\widetilde{U}(p_\shortparallel; y_3,x_3) \gamma_\mu
	\widetilde{U}(p_\shortparallel + k_\shortparallel; x_3,y_3) \gamma_\nu \right] \;.
\end{equation}
Inserting the explicit expressions for $\widetilde{S}_F^{(0)}$ and
$\widetilde{U}$ presented in the Appendix, one sees, after evaluating the
traces,  that $\Pi^{LU} + \Pi^{UL}$ vanishes identically. 
Thus, 
\begin{equation}\label{eq:pilu}	
\widetilde{\Pi}_{\mu\nu}(k_\shortparallel; x_3,y_3) \;=\; 
\widetilde{\Pi}^L_{\mu\nu}(k_\shortparallel; x_3,y_3) \,+\, 
\widetilde{\Pi}^U_{\mu\nu}(k_\shortparallel; x_3,y_3) \;, 
\end{equation}
where: $\widetilde{\Pi}^L_{\mu\nu}\equiv\widetilde{\Pi}^{LL}_{\mu\nu}$ and 
$\widetilde{\Pi}^U_{\mu\nu}\equiv\widetilde{\Pi}^{UU}_{\mu\nu}$. 

Before evaluating the above for some particular cases, we calculate a
magnitude corresponding to a related effect: the boundary conditions at $0$
and $\epsilon$ do break the chiral symmetry of the (massless) unconfined
theory. A quantitative and local measure of that violation, based on the
expectation value of a bilinear observable is the fermion condensate $\rho(x)
\equiv \langle {\bar\psi}(x) \psi(x)\rangle$.

In terms of the fermion propagator, we see that $\rho$ may be expressed as
follows:
\begin{equation}
\rho(x) \;=\; - {\rm tr}\big[ S_F(x,x) \big] \;.
\end{equation}
Furthermore, taking into account translation invariance in the parallel
coordinates, and the specific form of $\widetilde{S}_F$:
 \begin{equation}
\rho(x) \;=\; \rho(x_3) \;=\; - \int \frac{d^3p_\shortparallel}{(2\pi)^3} \; 
{\rm tr}\big[ \widetilde{S}_F(p_\shortparallel; x_3,x_3) \big] \;,
\end{equation}
which may be exactly calculated:
\begin{equation}
\rho(x_3) \;=\; - \frac{\pi^4}{2 \epsilon^3} \, \frac{3 + \cos(\frac{2\pi
x_3}{\epsilon})}{\sin^3(\frac{\pi x_3}{\epsilon})}\;,
\end{equation}
which is of course symmetric with respect to $x_3 = \frac{\epsilon}{2}$,
and finite everywhere, except at $x_3=0, \epsilon$. 
\begin{figure}
\begin{center}
\begin{picture}(0,0)%
\includegraphics{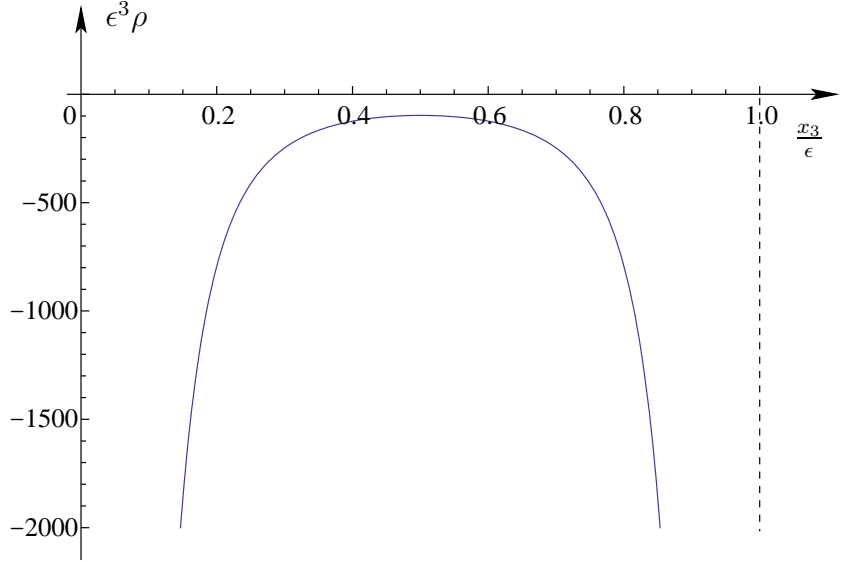}%
\end{picture}%
\setlength{\unitlength}{3947sp}%
\begingroup\makeatletter\ifx\SetFigFont\undefined%
\gdef\SetFigFont#1#2#3#4#5{%
  \reset@font\fontsize{#1}{#2pt}%
  \fontfamily{#3}\fontseries{#4}\fontshape{#5}%
  \selectfont}%
\fi\endgroup%
\begin{picture}(5187,3528)(-14,-12493)
\put(601,-9136){\makebox(0,0)[lb]{\smash{{\SetFigFont{12}{14.4}{\rmdefault}{\mddefault}{\updefault}{\color[rgb]{0,0,0}$\epsilon ^ 3 \rho$}%
}}}}
\put(4876,-9886){\makebox(0,0)[lb]{\smash{{\SetFigFont{12}{14.4}{\rmdefault}{\mddefault}{\updefault}{\color[rgb]{0,0,0}$\frac{x_3}{\epsilon}$}%
}}}}
\end{picture}%
\caption{$\epsilon^3 \rho$ as a function of $x_3 / \epsilon$.}
\end{center}
\end{figure}

As it can be inferred from  Figure 1, for finite values of $\epsilon$, $\rho(x_3)$ diverges on the boundaries, and has a maximum of $-\pi ^4/\epsilon ^3$ when $x_3 = \epsilon / 2$. It can also be shown that it tends to $0$ when $\epsilon
\to \infty$ and $x_3$ is far from the borders.
This concentration around the boundaries does also show up in the
calculation of the induced vacuum current, $j_\mu$, resulting from some
special electromagnetic field configurations. Those induced currents are,
in this linear response approximation, determined by $\Pi_{\mu\nu}$ and the
gauge field $A_\mu$ corresponding to the given electromagnetic field.
In our conventions:
\begin{eqnarray}
j_\mu(x) &=& e \, \langle {\bar\psi}(x) \gamma_\mu \psi(x) \rangle \nonumber\\
&=& - i \, \int d^4y \, \Pi_{\mu\nu}(x,y) \, A_\nu(y) \;,
\end{eqnarray}
and in the Fourier representation:
\begin{equation} 
\widetilde{j}_\mu(k_\shortparallel;x_3) \;=\; -i \int  dy_3 \,
\widetilde{\Pi}_{\mu\nu}(k_\shortparallel; x_3,y_3) \,
\widetilde{A}_\nu(k_\parallel; y_3) \;. 
\end{equation}
We have evaluated the current for the particular case
of a static electric field with normal incidence ($x_3$ direction).
This corresponds to $F_{01} = E$, where $E$ is a constant. In the $A_3=0$ gauge: 
$A_0 = - E x_3$, and $A_\mu=0$ for $\mu \neq 0$.
Thus,
\begin{equation} 
j_\mu = j_\mu(x_3) \;=\; i E  \int  dy_3 \,
\widetilde{\Pi}_{\mu 0}(0; x_3,y_3) \,y_3 \;, 
\end{equation}
and its usually more convenient real-time (Minkowski) version, $j^\mu_M$, becomes:
\begin{equation} 
j^0_M(x_3) \;=\; -  E  \int  dy_3 \, \widetilde{\Pi}_{0 0}(0; x_3,y_3) \,y_3 \;, 
\end{equation}
and, (with $l \neq 0$):
\begin{equation} 
j^l_M(x_3) \;=\; +  E  \int  dy_3 \, \widetilde{\Pi}_{l 0}(0; x_3,y_3) \,y_3 \;. 
\end{equation}
Since the system is parity conserving, $j_1\, = \, j_2=0$ for this external
field. Thus, we
concentrate on $j^0_M$ and $j^3_M$. Recalling (\ref{eq:pilu}), we see that
these currents receive two contributions:
\begin{eqnarray} 
j^0_M(x_3) &=& -  E  \int  dy_3 \, \Big[\widetilde{\Pi}^L_{0 0}(0; x_3,y_3)
\,y_3 \nonumber\\
 &+& \widetilde{\Pi}^U_{0 0}(0; x_3,y_3)\Big] \,y_3  \;.
\end{eqnarray}
and analogously for $j^3_M$.

To otain $j^0_M$, we note first that:
\begin{equation}
\widetilde{\Pi}_{00}^{L}(0; x_3, y_3) \;=\;-\frac{e^2}{(2\pi)^3} \, \int d^3p
\; e^{2\vert p \vert \, \vert x_3 - y_3 \vert } \,2\left( 1 -
\frac{p_0^{2}}{p^{2}}\right)\,=\,-\frac{e^2}{6\pi^2 \vert x_3 - y_3 \vert
^{3}}\;,
\end{equation}
which is finite for all $x_3 \neq y_3$. A divergence appear when
integrating
over $y_3$, to calculate the current. However, the usual renormalization
conditions imply that one has to use Hadamard's finite part in that
integral.

On the other hand, for $\widetilde{\Pi}_{00}^{U}$ we find the expression:
\begin{equation}
\widetilde{\Pi}_{00}^{U}(0; x_3, y_3) = \mathcal{M}_1\Big( \frac{x_3 - y_3 }{\epsilon}\Big) + \mathcal{M}_2\Big( \frac{x_3 + y_3 }{\epsilon}\Big)\;,
\end{equation}
where
\begin{eqnarray}
\mathcal{M}_1(u) &=& 
- \frac{e^2}{\epsilon^3} \left\lbrace -\frac{1}{72} + \frac{\zeta
  (3)}{8\pi^{2} } + F_1 (u)
\right\rbrace \bigg] \nonumber\\ 
\mathcal{M}_2(v) &=& 
- \frac{e^2}{\epsilon^3} \bigg[ \frac{5}{72} + F_2(v)  \bigg]
\end{eqnarray}
with
\begin{eqnarray}
F_1(u) &=& \frac{1}{24 \pi ^2 u^2} -\frac{\cos (\pi  u)}{24 \, \sin ^2  (\pi  u)} 
+ \frac{1}{384 \pi ^2} \bigg[(1-u) \psi
^{(2)}\left(\frac{1}{2}-\frac{u}{2}\right) \nonumber\\
&-& (1-u) \psi ^{(2)}\left(1-\frac{u}{2}\right) - (1+u) \psi ^{(2)}\left(1+\frac{u}{2}\right) 
+ (1+u) \psi ^{(2)}\left(\frac{1+u}{2}\right)
\bigg]  \nonumber\\
F_2(v) &=& \frac{1}{48} \csc ^3(\pi  v) \bigg[2 \sin (2 \pi  v)+\pi  (1-v) \Big(3+\cos (2 \pi  v)\Big)\bigg]\;.
\end{eqnarray} 
where $\zeta $ is Riemann's zeta function and $\psi ^{(2)}(z) =
\dfrac{d^3(\ln\Gamma(z))}{dz ^3}$.

\begin{figure}
\begin{center}
\begin{picture}(0,0)%
\includegraphics{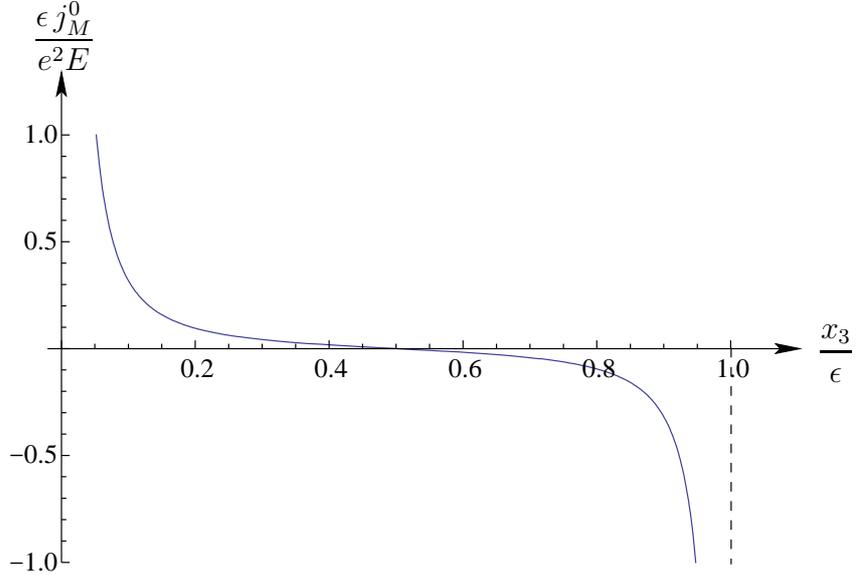}
\end{picture}%
\setlength{\unitlength}{3947sp}%
\begingroup\makeatletter\ifx\SetFigFont\undefined%
\gdef\SetFigFont#1#2#3#4#5{%
  \reset@font\fontsize{#1}{#2pt}%
  \fontfamily{#3}\fontseries{#4}\fontshape{#5}%
  \selectfont}%
\fi\endgroup%
\begin{picture}(5055,3469)(-14,-12269)
\put(5026,-10936){\makebox(0,0)[lb]{\smash{{\SetFigFont{12}{14.4}{\rmdefault}{\mddefault}{\updefault}{\color[rgb]{0,0,0}$\dfrac{x_3}{\epsilon}$}%
}}}}
\put(148,-8983){\makebox(0,0)[lb]{\smash{{\SetFigFont{12}{14.4}{\rmdefault}{\mddefault}{\updefault}{\color[rgb]{0,0,0}$\dfrac{\epsilon \,j^0_M }{e^2 E} $}%
}}}}
\end{picture}%
\caption{$\dfrac{\epsilon\, j^0_M}{e^2 E}  $ as a function of $\dfrac{x_3}{\epsilon}$.}\label{fg:2} 
\end{center}
\end{figure}

In Figure 2, we plot the profile of $\dfrac{\epsilon\, j^0_M}{e^2 E}  $ as a function of $\dfrac{x_3}{\epsilon}$ that follows from the previous results.

For $j^3_M$, under the same external field configuration, we have found a
vanishing result, namely, $j^3_M=0$. It should not be surprising that the
normal current vanishes at the borders, since the bag conditions should do
precisely that. The vanishing of the current inside the slab, however, should be regarded as a
steady state configuration feature, reached after the charge density has
adopted the profile above. 

The profile for the charge density above corresponds to the case of a
normal electric field; to explore a situation that is, somehow, opposite
to that one, we presenting here an exact expression for the effective action,
$\Gamma_f$, in the case where the gauge field depends only on the parallel
coordinates, and moreover that field is a function of only the
parallel coordinates. Following the approach of~\cite{EfeCasFer} for the case at
hand, an entirely analogous procedure to the one used there, yields now:
\begin{equation}\label{eq:gammaef}
e^{-\Gamma_f(A_\parallel)} \;=\; \det {\mathcal K}
\end{equation}
where ${\mathcal K}$ is the object
\begin{equation}
{\mathcal K}=\left[\begin{array}{cc}
      1 + V & (V -\gamma_3 ) 
     e^{- \epsilon {\mathcal H} } \\
                    \\
     (V +\gamma_3) e^{- \epsilon {\mathcal H}}& 
     1 + V  \\
\end{array}\right]
\end{equation}
with 
\begin{equation}
{\mathcal H} \equiv  \sqrt{-\not \!\! D_\parallel^2} \;,\;\;\;  V \equiv
\frac{- \not \!\! D_\parallel}{\mathcal H} \;.
\end{equation}
Here, $\not\!\!\!D_\parallel$ is the `parallel Dirac operator': 
$\not \!\!\!D_\parallel = \gamma_\alpha D_\alpha$, $\alpha=0,1,2$, and
$D=\partial_\parallel + A_\parallel(x_\parallel)$.   

Using some algebra, the determinant above may also be written in the equivalent form:
\begin{equation}
 \det {\mathcal K} \;=\; \big[\det (1 + V)\big]^2 \, \det\Big[ 1 - \frac{1 +
\gamma_3}{2} ( 1 + V) e^{- 2 \epsilon {\mathcal H}} \Big] \;. 
\end{equation}
As a consequence, $\Gamma_f(A_\parallel)$ receives two contributions, one
coming from the determinant of $(1 + V)$, which is essentially two
dimensional and independent of $\epsilon$, plus an extra term
$\Gamma_\epsilon$, which does depend on $\epsilon$, it is finite (because of the exponential factor),
and tends to zero when $\epsilon \to \infty$:
\begin{equation}\label{eq:gammae}
\Gamma_\epsilon(A_\parallel) \;=\; - {\rm Tr} \log \left[ 1 - \frac{1 +
\gamma_3}{2} ( 1 + V) e^{- 2 \epsilon {\mathcal H}} \right] \;.
\end{equation}

We end this note by presenting our conclusions: We have found the form of
of the induced vacuum current due to a normal electric field, which shows that the
induced charge density distributes itself in order to counterbalance the
external electric field. We obtained an expression for the effective action
that follows from considering a parallel gauge field configuration. It
contains a term which represents the finite-width contribution to the
effective action. It is finite, and it contains a  $\frac{1 +
\gamma_3}{2}$ factor, which is a projector which accounts for the suppression of part of the
fermion modes because of the boundary conditions. 
\section*{Acknowledgements}
The authors thank CONICET and UNCuyo for financial support.


\begin{thebibliography}{bib}
\bibitem{Casimir:1948dh}
H.~B.~G.~Casimir,
Indag.\ Math.\  {\bf 10}, 261 (1948) [Kon.\ Ned.\ Akad.\ Wetensch.\
Proc.\  {\bf 51}, 793 (1948\ FRPHA,65,342-344.1987\
KNAWA,100N3-4,61-63.1997)].
\bibitem{Bordag:2001qi}
For a review, see, for example: M.~Bordag, U.~Mohideen and V.~M.~Mostepanenko,
Phys.\ Rept.\  {\bf 353}, 1 (2001).
\bibitem{Chodos:1974pn}
A.~Chodos, R.~L.~Jaffe, K.~Johnson and C.~B.~Thorn,
Phys.\ Rev.\  D {\bf 10}, 2599 (1974).
\bibitem{Milton:2001yy}
K.~A.~Milton,
{\it  River Edge, USA: World Scientific (2001) 301 p}
\bibitem{EfeCasFer}
C.D. Fosco and E. Losada; Phys.Rev.D, 78, 025017, (2008).- arXiv:0805.2922 [hep-th].
\bibitem{Balian:1970fw}
R.~Balian and C.~Bloch, 
Annals Phys.\  {\bf 60}, 401 (1970);
Ibid., Annals Phys.\  {\bf 64}, 271 (1971).
\bibitem{Hansson:1982cu}
  T.~H.~Hansson and R.~L.~Jaffe,
  Phys.\ Rev.\  D {\bf 28}, 882 (1983).
\bibitem{Hansson:1983xt}
  T.~H.~Hansson and R.~L.~Jaffe,
   ``The Multiple Reflection Expansion For Confined Scalar, Dirac And Gauge
  Annals Phys.\  {\bf 151}, 204 (1983).
\bibitem{Fosco:2004cn}
  C.~D.~Fosco, A.~P.~C.~Malbouisson and I.~Roditi,
  Phys.\ Lett.\  B {\bf 609}, 430 (2005)
  [arXiv:hep-th/0412229].
\end{thebibliography}
\end{document}